\documentclass[12pt]{iopart}

\usepackage{iopams}
\usepackage{moreverb,url}
\newcommand{\emp}[1]{{\texttt{#1}}}

\newcommand{\matenv}[1]{${#1}$}
\newcommand{\oper}[1]{ #1}
\newcommand{\refeq}[1]{(\ref{#1})}
\renewcommand{\vec}[1]{\boldsymbol{\mathbf{#1}}}
\newtheorem{algorithm}{Algorithm}{\bfseries}{\rm}

\begin{document}

\title[Simplifying Parallelization of Scientific Codes]{Simplifying
  Parallelization of Scientific Codes by a Function-Centric Approach
  in Python}

\author{Jon K. Nilsen$^{1,2}$, Xing Cai$^{3,4}$, Bj{\o}rn
    H{\o}yland$^{5}$ and Hans Petter Langtangen$^{3,4}$}

\address{ $^1$ USIT, P.O.~Box 1059 Blindern, N-0316 Oslo,
    Norway\\ $^2$ Department of Physics, P.O.~Box 1048 Blindern,
    University of Oslo, N-0316 Oslo, Norway\\ $^3$ Center for
    Biomedical Computing, Simula Research Laboratory, P.O.~Box 134,
    N-1325 Lysaker, Norway\\ $^4$ Department of Informatics, P.O.~Box
    1080 Blindern, University of Oslo, N-0316 Oslo, Norway\\ $^5$
    Department of Political Science, P.O.~Box 1097 Blindern,
    University of Oslo, N-0317 Oslo, Norway}
\eads{\mailto{j.k.nilsen@usit.uio.no}, \mailto{xingca@simula.no},
  \mailto{bjorn.hoyland@stv.uio.no}, \mailto{hpl@simula.no}}

\begin{abstract}

The purpose of this paper is to show how existing scientific software
can be parallelized using a separate thin layer of Python code where
all parallel communication is implemented. We provide specific
examples on such layers of code, and these examples may act as
templates for parallelizing a wide set of serial scientific codes.
The use of Python for parallelization is motivated by the fact that
the language is well suited for reusing existing serial codes
programmed in other languages. The extreme flexibility of Python with
regard to handling functions makes it very easy to wrap up decomposed
computational tasks of a serial scientific application as Python
functions.  Many parallelization-specific components can be
implemented as generic Python functions, which may take as input those
functions that perform concrete computational tasks.  The overall
programming effort needed by this parallelization approach is rather
limited, and the resulting parallel Python scripts have a compact and
clean structure.  The usefulness of the parallelization approach is
exemplified by three different classes of applications in natural and
social sciences.

\end{abstract}

\submitto{Computational Science \& Discovery}

\maketitle

\section{Introduction}
Due to limited computing power of standard serial computers, parallel
computing has become indispensable for investigating complex problems
in all fields of science. A frequently encountered question is how to
transform an existing serial scientific code into a new form that is
executable on a parallel computing platform. Although portable
parallel programming standards, such as MPI and OpenMP, have greatly
simplified the programming work, the task of parallelization may still
be quite complicated for domain scientists. This is because inserting
MPI calls or OpenMP directives directly into an existing serial code
often requires extensive code rewrite as well as detailed knowledge of
and experience with parallel programming.

The hope for non-specialists in parallel computing is that many
scientific applications possess high-level parallelism.  That is, the
entire computational work can be decomposed into a set of individual
(and often collaborative) computational tasks, each of coarse grain,
and can be performed by an existing piece of serial code.  Depending
on the specific application, the decomposition can be achieved by
identifying a set of different parameter combinations, or (fully or
almost) independent computations, or different data groups, or
different geometric subdomains.  For a given type of decomposition,
the parallelization induced programming components, such as work
partitioning, domain partitioning, communication, load balancing, and
global administration, are often generic and independent of specific
applications.  These generic components can thus be implemented as
reusable parallelization libraries once and for all.  This is what we
exemplify in the present paper.

It is clear that a user-friendly parallelization approach relies on at
least two factors: (1) The existing serial code should be extensively
reused; (2) The programming effort by the end user must be limited.
To achieve these goals we suggest to use Python to wrap up pieces of
existing serial code (possibly written in other languages), and
implement the parallelization tasks in separate and generic Python
functions.

Python~\cite{python} is an extremely expressive and flexible
programming language at its core. The language has been extended with
numerous numerical and visualization modules such as
NumPy~\cite{NumPy} and SciPy~\cite{scipy}.  The two requirements of a
user-friendly parallelization mentioned above are actually well met by
Python.  First of all, Python is good at inter-operating with other
languages, especially Fortran, C, and C++, which are heavily used in
scientific codes.  Using wrapper tools such as F2PY~\cite{f2py}, it is
easy to wrap up an existing piece of code in Fortran and C and provide
it with a Pythonic appearance.

Moreover, among its many strong features, Python is extremely flexible
with handling functions.  Python functions accept both positional
arguments and keyword arguments.  The syntax of a variable set of
positional and keyword arguments (known as ``\emp{(*args,**kwargs)}''
to Python programmers) allows writing libraries routines that work
with any type of user-defined functions.  That is, the syntax makes it
possible to call a Python function without revealing the exact number
of arguments.

It is also straightforward to pass functions as input arguments to a
Python function and/or return a function as output.  A callable class
object in Python can be used as if it were a stand-alone
function. Such a construction, or alternatively a closure (known from
functional programming), can be used to create functions that carry a
state represented through an arbitrarily complex data structure. The
result is that one can express the flow of a scientific code as a
Python program containing a set of calls to user-defined Python
functions. These user-defined functions can be ordinary functions or
classes that wrap pieces of the underlying scientific code.  This is
what we call a \emph{function-centric} representation of the
scientific code. With such a function-centric approach, we can build a
general framework in Python for almost automatic parallelization of
the program flow in the original scientific code.  Later examples will
convey this idea in detail.

Performance of the resulting parallel application will closely follow
the performance of the serial application, because the overhead of the
parallelization layer in Python is just due to a small piece of extra
code, as we assume the main computational work to take place in the
Python functions that call up pieces of the original scientific code.
In the parallelization layer, good performance can be ensured by using
efficient array modules in Python (such as \emp{numpy}~\cite{NumPy})
together with light-weight MPI wrappers (such as
\emp{pypar}~\cite{Pypar}).  For examples of writing efficient Python
code segments for some standard serial and parallel numerical
computations, we refer the reader to Cai et al.~\cite{Cai2005SP}.

\paragraph{Related Work.}
In C++, generic programming via templates and object-oriented
programming has been applied to parallelizing serial scientific codes.
Two examples can be found in~\cite{SC.5.Cai.2002.a}
and~\cite{Cirak2008}, where the former uses C++ class hierarchies to
enable easy implementation of additive Schwarz preconditioners, and
the latter uses C++ templates extensively to parallelize finite
element codes.  Many scientific computing frameworks have also adopted
advanced programming to incorporate parallelism behind the scene. In
these frameworks (see, e.g.,
\cite{petsc-web-page,ParFUM,Sierra,Trilinos,UG}) the users can write
parallel applications in a style quite similar to serial programming,
without being exposed to many parallelizing details. Likewise are
frameworks that are specially designed to allow coupling of different
serial and parallel components, such as Cactus~\cite{Cactus} and
MpCCI~\cite{MpCCI}.  The Python programming language, however, has not
been widely used to parallelize existing serial codes. The Star-P
system~\cite{Star-P} provides the user with a programming environment
where most of the parallelism is kept behind the
scene. Hinsen~\cite{Hinsen07} has combined Python with BSP to enable
high-level parallel programming. In addition, quite a number of Python
MPI wrappers exist, such as \emp{pyMPI}~\cite{pyMPI},
\emp{pypar}~\cite{Pypar}, \emp{MYMPI}~\cite{myMPI},
\emp{mpi4py}~\cite{mpi4python,mpi4python2}, and
\emp{Scientific.MPI}~\cite{Scientific}.  Efforts in incorporating
parallelism via language extensions of Python can be found
in~\cite{Benson07,cp-stackless-python,python-coarray}.

The contribution of the present paper is to show by examples that a
function-centric approach using Python may ease the task of parallel
scientific programming.  This result is primarily due to Python's
flexibility in function handling and function arguments.  As a result,
generic tasks that arise in connection with parallelization can often
be programmed as a collection of simple and widely applicable Python
functions, which are ready to be used by non-specialists to
parallelize their existing serial codes.

This paper contains three examples with different algorithmic
structures. A wide range of problems in science can be attacked by
extending and adapting the program code in these examples.  Moreover,
readers whose problems are not covered by the examples will hopefully
from these examples understand how we solve programming problems by
identifying the principal, often simplified, underlying algorithmic
structure; then creating generic code to reflect the structure; and
finally applying the generic code to a specific, detailed case.  Our
approach is much inspired by the success of mathematics in problem
solving, i.e., detecting the problem's principal structure and
devising a generic solution makes complicated problems tractable.
With Python as tool, we demonstrate how this strategy carries over to
parallelization of scientific codes.

The remainder of the paper is organized as follows.  We give in
Section~\ref{sect:motive} a simple but motivating example, explaining
the principles of splitting a problem into a set of function calls
that can easily be parallelized.  Generic parallelization of three
common types of real scientific applications are then demonstrated in
Section~\ref{sect:highlevel}.  Afterwards, Section~\ref{sect:results}
reports the computational efficiency of the suggested parallelization
approach applied to specific cases in the three classes of scientific
problems.  Some concluding remarks are given in
Section~\ref{sect:conclusion}.

\section{A Motivating Simple Example}
\label{sect:motive}

\subsection{Serial Version}
\label{sect:motive:serial}

Suppose we want to carry out a parameter analysis that involves a
large number of evaluations of a multi-variable mathematical function
$f(a_1,\ldots,a_q)$.  The Python implementation of $f$ may use $p$
positional arguments and $k$ keyword arguments such that the total
$p+k$ arguments contain at least the variables $a_1,\ldots,a_q$ (i.e.,
$q\leq p+k$).  As a very simple example, consider the parabola
$f(x,a,b,c)=ax^2+bx+c$ with the following Python implementation ($q=4,
p=1, k=3$): {
\begin{quote}
\renewcommand{\baselinestretch}{0.85}\footnotesize
\begin{verbatimtab}[8]
def func(x, a=0, b=0, c=1):
    return a*x**2+b*x+c
\end{verbatimtab}
\end{quote}
}
\noindent
Suppose we want to evaluate \emp{func} for a particular set of input
parameters chosen from a large search space, where $x$, $a$, $b$, and
$c$ vary in specified intervals.  The complete problem can be
decomposed into three main steps: (1) initialize a set of arguments to
\emp{func}; (2) evaluate \emp{func} for each entry in the set of
arguments; (3) process the set of function return values from all the
\emp{func} calls.

Step (1) calls a user-defined function \emp{initialize} which returns
a list of 2-tuples, where each 2-tuple holds the positional and
keyword arguments (as a tuple and a dictionary) for a specific call to
\emp{func}. Step (2) iterates over the list from step (1) and feed the
positional and keyword arguments into \emp{func}. The returned value
(tuple) is stored in a result list. Finally, step (3) processes the
result list in a user-defined function \emp{finalize} which takes this
list as input argument.

A generic Python function that implements the three-step parameter
analysis can be as follows: {
\begin{quote}
\renewcommand{\baselinestretch}{0.85}\footnotesize
\begin{verbatimtab}[8]
def solve_problem(initialize, func, finalize):
    input_args = initialize()
    output = [func(*args, **kwargs) for args, kwargs in input_args]
    finalize(output)
\end{verbatimtab}
\end{quote}
} Note that the use of {\em list comprehension} in the above code has
given a very compact implementation of the \emp{for}-loop for going
through all the evaluations of \emp{func}.  The \emp{initialize},
\emp{func}, and \emp{finalize} functions are passed to
\emp{solve\_problem} as input arguments. These three user-defined
functions are independent of \emp{solve\_problem}.

As an example, assume that \emp{x} is a set of $n$ uniformly
distributed coordinates in $[0,L]$, and we vary $a$ and $b$ in
$[-1,1]$ each with $m$ values, while $c$ is fixed at the value 5. For
each combination of $a$ and $b$, we call \emp{func} with the vector
\emp{x} as a positional argument and the $a$, $b$, $c$ values as
keyword arguments, and store the evaluation results of \emp{func} in a
list named \emp{output}. The objective of the computations is to
extract the $a$ and $b$ values for which \emp{func} gives a negative
value for one or several of the coordinates $x\in [0,L]$.  For this
very simple example, the concrete implementation of the
\emp{initialize} and \emp{finalize} functions can be put inside a
class named \emp{Parabola} as follows: {
\begin{quote}
\renewcommand{\baselinestretch}{0.85}\footnotesize
\begin{verbatimtab}[8]
class Parabola:
    def __init__(self, m, n, L):
        self.m, self.n, self.L = m, n, L

    def initialize(self):
        x = numpy.linspace(0, self.L, self.n)
        a_values = numpy.linspace(-1, 1, self.m)
        b_values = numpy.linspace(-1, 1, self.m)
        c = 5

        self.input_args = []
        for a in a_values:
            for b in b_values:
                func_args = ([x], {'a': a, 'b': b, 'c': c})
                self.input_args.append(func_args)
        return self.input_args

    def func(self, x, a=0, b=0, c=1):
        return a*x**2+b*x+c

    def finalize(self, output_list):
        self.ab = []
        for input, result in zip(self.input_args, output_list):
            if min(result) < 0:
                self.ab.append((input[0][1], input[0][2]))
\end{verbatimtab}
\end{quote}
}

Now, to find the combinations of $a$ and $b$ values that make
$ax^2+bx+c<9$, we can write the following two lines of code (assuming
$m=100$, $n=50$, and $L=10$): {
\begin{quote}
\renewcommand{\baselinestretch}{0.85}\footnotesize
\begin{verbatimtab}[8]
problem = Parabola(100, 50, 10)
solve_problem(problem.initialize, problem.func, problem.finalize)
\end{verbatimtab}
\end{quote}
}
\noindent Note that the desired combinations of $a$ and $b$ values
will be stored in the list \emp{problem.ab}.  Also note that we have
placed \emp{func} inside class \emp{Parabola}, to have all pieces of
the problem in one place, but having \emp{func} as stand-alone
function or a class method is a matter of taste.

Despite the great mathematical simplicity of this example, the
structure of the \emp{solve\_problem} function is directly applicable
to a wide range of much more advanced problems.  Although
\emp{initialize} and \emp{finalize} are Python functions with very
simple arguments (none and a list, respectively), this is not a
limitation of their applicability.  For example, the \emp{initialize}
step in our simple example needs values for $m$, $n$, and $L$, the $a$
and $b$ interval and so on, which can not be specified in the generic
\emp{solve\_problem} function.  To overcome this limitation, the
information of $m$, $n$, and $L$ can be hard-coded (not recommended),
or transferred to \emp{initialize} through global variables (not
recommended in general) or carried with \emp{initialize} as a state,
either as class attributes or as a surrounding scope in a closure. We
have chosen the class approach, i.e., class attributes store
user-dependent data structures such that the \emp{initialize} and
\emp{finalize} methods can have the simple input argument structure
demanded by the generic \emp{solve\_problem} function. Alternatively,
a closure as follows can be used instead of a class (this construct
requires some knowledge of Python's scoping rules): {
\begin{quote}
\renewcommand{\baselinestretch}{0.85}\footnotesize
\begin{verbatimtab}[8]
def initialize_wrapper(m, n, L):
    def initialize(self):
        x = numpy.linspace(0, L, n)
        a_values = numpy.linspace(-1, 1, m)
        ...
        return input_args
    return initialize
\end{verbatimtab}
\end{quote}
} Now, the returned \emp{initialize} function will carry with it the
values of $m$, $n$, and $L$ in the surrounding scope. The choice
between the class approach and the closure approach, or using global
variables in a straightforward global \emp{initialize} function, is up
to the programmer.  The important point here is that \emp{initialize}
must often do a lot, and the input information to \emp{initialize}
must be handled by some Python construction. Similar comments apply to
\emp{finalize}.

\subsection{Parallel Version}
\label{sect:motive:parallel}
Let us say that we want to utilize several processors to share the
work of all the \emp{func} evaluations, i.e., the \emp{for}-loop in
the generic \emp{solve\_problem} function.  This can clearly be
achieved by a task-parallel approach, where each evaluation of
\emp{func} is an independent task.  The main idea of parallelization
is to split up the \emp{for}-loop into a set of shorter
\emp{for}-loops, each assigned to a different processor.  In other
words, we need to split up the \emp{input\_args} list into a set of
sub-lists for the different processors.  Note that this partitioning
work is generic, independent of both the \emp{func} function and the
actual arguments in the \emp{input\_args} list. Assuming homogeneous
processors and that all the function evaluations are equally
expensive, we can divide the \emp{input\_args} list into
\emp{num\_procs} (number of processors) sub-lists of equal length.  In
case \emp{input\_args} is not divisible by \emp{num\_procs}, we adjust
the length of some sub-lists by 1:
% code taken from function_centric.py
{
\begin{quote}
\renewcommand{\baselinestretch}{0.85}\footnotesize
\begin{verbatimtab}[8]
def simple_partitioning(length, num_procs):
    sublengths = [length/num_procs]*num_procs
    for i in range(length % num_procs):  # treatment of remainder
        sublengths[i] += 1
    return sublengths

def get_subproblem_input_args(input_args, my_rank, num_procs):
    sub_ns = simple_partitioning(len(input_args), num_procs)
    my_offset = sum(sub_ns[:my_rank])
    my_input_args = input_args[my_offset:my_offset+sub_ns[my_rank]]
    return my_input_args
\end{verbatimtab}
\end{quote}
} Using the above generic \emp{get\_subproblem\_input\_args} function,
each processor gets its portion of the global \emp{input\_args} list,
and a shorter \emp{for}-loop can be executed there.  Note that the
syntax of Python lists and \emp{numpy} arrays has made the function
very compact.

The next step of parallelization is to collect the function evaluation
results from all the processors into a single global \emp{output}
list.  Finally, we let \emp{finalize(output)} run only on the master
processor (assuming that this work does not require parallelization).
For the purpose of collecting outputs from all the processors, the
following generic Python function can be used:
% also taken from function_centric.py
{
\begin{quote}
\renewcommand{\baselinestretch}{0.85}\footnotesize
\begin{verbatimtab}[8]
def collect_subproblem_output_args(my_output_args, my_rank, 
                                   num_procs, send_func, recv_func):
    if my_rank == 0:    # master process?
        output_args = my_output_args
        for i in range(1, num_procs):
            output_args += recv_func(i)
        return output_args
    else:
        send_func(my_output_args, 0)
        return None
\end{verbatimtab}
\end{quote}
}

The last two input arguments to the above function deserve some
attention.  Both \emp{send\_func} and \emp{recv\_func} are functions
themselves.  In the case of using the \emp{pypar} wrapper of MPI
commands, we may simply pass \emp{pypar.send} as the \emp{send\_func}
input argument and \emp{pypar.receive} as \emp{recv\_func}. Moreover,
switching to another MPI module is transparent with regard to the
generic function named \emp{collect\_subproblem\_output\_args}.  It
should also be noted that most Python MPI modules are considerably
more user-friendly than the original MPI commands in C/Fortran.  This
is because (1) the use of keyword arguments greatly simplifies the
syntax, and (2) any picklable (marshalable) Python data type can be
communicated directly.

Now that we have implemented the generic functions
\emp{get\_subproblem\_input\_args} and
\emp{collect\_subproblem\_output\_args}, we can write a minimalistic
parallel solver as follows: {
\begin{quote}
\renewcommand{\baselinestretch}{0.85}\footnotesize
\begin{verbatimtab}[8]
def parallel_solve_problem(initialize, func, finalize,
                           my_rank, num_procs, send, recv):
    input_args = initialize()
    my_input_args = get_subproblem_input_args(input_args,
                                              my_rank, num_procs)
    my_output = [func(*args, **kwargs) \
                 for args, kwargs in my_input_args]
    output = collect_subproblem_output_args(my_output, my_rank, 
                                            num_procs, send, recv)
    if my_rank == 0:
        finalize(output)
\end{verbatimtab}
\end{quote}
}We remark that the above function is generic in the sense that it is
independent of the actual implementation of \emp{initialize},
\emp{func}, and \emp{finalize}, as well as the Python MPI module being
used.  All problems that can be composed from independent function
calls can (at least in principle) be parallelized by the shown small
pieces of Python code.

As a specific example of using this parallel solver, we may address
the problem of evaluating the parabolic function (\emp{func} and class
\emp{Parabola}) for a large number of parameters. Using the
\emp{pypar} MPI module and having the problem-dependent code in a
module named \emp{Parabola} and the general function-centric tools in
a module named \emp{function\_centric}, the program becomes as
follows:
% parabola_problem.py
{
\begin{quote}
\renewcommand{\baselinestretch}{0.85}\footnotesize
\begin{verbatimtab}[8]
from Parabola import func, Parabola
from function_centric import parallel_solve_problem

problem = Parabola(m=100, n=50, L=10)
import pypar
my_rank = pypar.rank()
num_procs = pypar.size()
parallel_solve_problem(problem.initialize,
                       func,
                       problem.finalize,
                       my_rank, num_procs,
                       pypar.send, pypar.receive)
pypar.finalize()
\end{verbatimtab}
\end{quote}
} To the reader, it should be obvious from this generic example how to
parallelize other independent function calls by the described
function-centric approach.

\section{Function-Centric Parallelization}
\label{sect:highlevel}

We have shown how to parallelize a serial program that is decomposable
into three parts: \emp{initialize}, calls to \emp{func} (i.e., a set
of independent tasks), and \emp{finalize}.  In this section, we
describe how the function-centric parallelization is helpful for three
important classes of scientific applications: Markov chain Monte Carlo
simulations, dynamic population Monte Carlo simulations, and solution
of partial differential equations.  We use Python to program a set of
simple and generic parallelization functions.

\subsection{Parallel Markov chain Monte Carlo Simulations}
\label{sect:3.1}

The standard Markov chain Monte Carlo algorithms are embarrassingly
parallel and have exactly the {\em same} algorithmic structure as the
example of parameter analysis in Section~\ref{sect:motive}.  This
means that the functions \emp{initialize}, \emp{func}, and
\emp{finalize} can easily be adapted to Monte Carlo problems.  More
specifically, the \emp{initialize} function prepares the set of random
samples and other input parameters. Some parametric model is computed
by the \emp{func} function, whereas \emp{finalize} collects the data
returned from all the \emp{func} calls and prepares for further
statistical analysis.

Function-centric parallelization of Markov chain Monte Carlo
applications closely follows the example in Section~\ref{sect:motive}.
We can reuse the three generic functions named
\emp{get\_subproblem\_input\_args},
\emp{collect\_subproblem\_output\_args}, and
\emp{parallel\_solve\_problem}, assuming that all the \emp{func}
evaluations are equally costly and all the processors are equally
powerful so there is no need for more sophisticated load balancing.

In Section~\ref{sect:pMCMC}, we will look at a real-life Markov chain
problem from political science (Appendix~\ref{sect:voting} gives its
mathematical description).

\subsection{Population Monte Carlo with Dynamic Load Balancing}
\label{sect:3.2}

A more advanced branch of Monte Carlo algorithms is population Monte
Carlo, see~\cite{iba-2000-16}. Here, a group of walkers, also called
the {\em population}, is used to represent a high-dimensional vector
and the computation is carried out by a random walk in the state
space. During the computation some of these walkers may be duplicated
or deleted according to some acceptance/rejection criteria, i.e., the
population is dynamic in time. Population Monte Carlo algorithms have
been proven useful in a number of fields, spanning from polymer
science to statistical sciences, statistical physics, and quantum
physics.

Unlike the examples so far, where the computational tasks were totally
independent and of static size, population Monte Carlo algorithms may
be viewed as an iteration in time where we repeatedly do some work on
a dynamic population, including moving the walkers of the population
and adjusting the population size, which in a parallel context calls
for dynamic load balancing.

\subsubsection{Serial Implementation}

A serial implementation of the time integration function can be as follows:
{
\begin{quote}
\renewcommand{\baselinestretch}{0.85}\footnotesize
\begin{verbatimtab}[8]
def time_integration(initialize, do_timestep, finalize):
    walkers, timesteps = initialize()
    output = []
    for step in range(timesteps):
        old_walkers_len = len(walkers)
        output.append(do_timestep(walkers))
        walkers.finalize_timestep(old_walkers_len, len(walkers))
    finalize(output)
\end{verbatimtab}
\end{quote}
}

The input arguments to the generic \emp{time\_integration} function
are three functions: \emp{initialize}, \emp{do\_timestep}, and
\emp{finalize}. This resembles the three-step structure discussed in
Section~\ref{sect:motive}.  The \emp{do\_timestep} function can have a
unified implementation for all the variants of population Monte Carlo
algorithms.  The other two input functions are typically programmed as
methods of a class that implements a particular algorithm (such as
diffusion Monte Carlo in Section~\ref{sect:pdmc}).  Here, the
\emp{initialize} method sets up a population object \emp{walkers} (to
be explained below) and determines the number of time steps the
walkers are to be propagated.  The \emp{finalize} method can, e.g.,
store the output for later analysis.

The purpose of the \emp{do\_timestep} function is to implement the
work for one time step, including propagating the walkers and
adjusting the population.  An implementation that is applicable for
all population Monte Carlo algorithms may have the following form:
\begin{quote}
{
\renewcommand{\baselinestretch}{0.85}\footnotesize
\begin{verbatimtab}[8]
def do_timestep(walkers):
    walkers.move()
    for walker in range(len(walkers)):
        if walkers.get_marker(walker) == 0:
            walkers.delete(walker)
        elif walkers.get_marker(walker) > 1:
            walkers.append(walker, walkers.get_marker(walker)-1)        
    return walkers.sample_observables()
\end{verbatimtab}
}
\end{quote}

The above implementation of \emp{time\_integration} and
\emp{do\_timestep} assumes that \emp{walkers} is an object of a class,
say with name \emp{Walkers}, that has a certain number of methods. Of
course, the flexibility of Python allows that the concrete
implementation of class \emp{Walkers} be made afterwards, unlike C++
and Java that require class \emp{Walkers} be written before
implementing \emp{time\_integration} and \emp{do\_timestep}.  Here, we
expect class \emp{Walkers} to provide a generic implementation of a
group of walkers, with supporting methods for manipulating the
population. The most important methods of class \emp{Walkers} are as
follows:
\begin{itemize}
\item \emp{move()} carries out the work of moving each walker of the
  population randomly according to some rule or distribution function.
\item \emp{get\_marker(walker)} returns the number of copies belonging
  to a walker with index \emp{walker}, where 0 means the walker should
  be deleted, 2 or more means that clones should be created.
\item \emp{append(walker, nchilds)} and \emp{delete(walker)} carry out
  the actual cloning and removal of a walker with index \emp{walker}.
\item \emp{sample\_observables()} returns the observables at a given
  time step, e.g., an estimate of the system energy.
\item \emp{finalize\_timestep(old\_size, new\_size)} does some
  internal book keeping at the end of each time step, such as
  adjusting some internal variables. It takes as input the total
  number of walkers before and after the walker population has been
  adjusted by the \emp{do\_timestep} function.
\item \emp{\_\_len\_\_} is one of Python's special class methods and
  is in our case meant to return the number of walkers.  A call
  \emp{len(walkers)} yields the same result as
  \emp{walkers.\_\_len\_\_()}.
\end{itemize}

For a real application, such as the diffusion Monte Carlo algorithm
(see Section~\ref{sect:pdmc} and Appendix~\ref{sect:dmc}), the
concrete implementation of the methods should reflect the desired
numerical algorithm.  For example, the \emp{move} method of diffusion
Monte Carlo uses diffusion and branching as the rule to randomly move
each walker, and the \emp{finalize\_timestep} method adjusts the
branching ratio.

\subsubsection{Parallelization}

Parallelism in population Monte Carlo algorithms arises naturally from
dividing the walkers among the processors.  Therefore, a parallel
version of the \emp{time\_integration} function may be as follows: {
\begin{quote}
\renewcommand{\baselinestretch}{0.85}\footnotesize
\begin{verbatimtab}[8]
def parallel_time_integration(initialize, do_timestep, finalize, 
                              my_rank, num_procs, send, recv, all_gather):
    my_walkers, timesteps = initialize(my_rank, num_procs)
    old_walkers_len = sum(all_gather(numpy.array([len(my_walkers)])))
    my_output = []
    for step in range(timesteps):
        # do what is required at this time step and measure CPU time
        t_start = time.time()
        results = do_timestep(my_walkers)
        my_output.append(results)
        task_time = time.time() - t_start

        # redistribute walkers and get walker size per process
        num_walkers_per_proc = dynamic_load_balancing(\
            my_walkers, task_time, my_rank, num_procs,\
            send, recv, all_gather)

        # finalize task for this time step
        new_walkers_len = sum(num_walkers_per_proc)
        my_walkers.finalize_timestep(old_walkers_len, new_walkers_len)
        old_walkers_len = new_walkers_len       
    my_output = collect_subproblem_output_args(my_output, my_rank,
                                               num_procs, send, recv)
    if my_rank == 0:
        finalize(my_output)
\end{verbatimtab}
\end{quote}
}

In comparison with its serial counterpart, the
\emp{parallel\_time\_integration} function has a few noticeable
changes.  First, the input arguments have been extended with five new
arguments. The two integers \emp{my\_rank} and \emp{num\_procs} are,
as before, meant for identifying the individual processors and finding
the total number of processors.  The other three new input arguments
are MPI communication wrapper functions: \emp{send}, \emp{recv}, and
\emp{all\_gather}, which can be provided by any of the Python wrapper
modules of MPI. The only exception is that \emp{pypar} does not
directly provide the \emp{all\_gather} function, but we can easily
program it as follows: {
\begin{quote}
\renewcommand{\baselinestretch}{0.85}\footnotesize
\begin{verbatimtab}[8]
    def all_gather (input_array):
          array_gathered_tmp = pypar.gather (input_array, 0)
          array_gathered = pypar.broadcast (array_gathered_tmp, 0)
          return array_gathered
\end{verbatimtab}
\end{quote}
}

Second, we note that the \emp{initialize} function is slightly
different from the serial case, now accepting \emp{my\_rank} and
\emp{num\_procs} as input. This is because initial division of the
walkers is assumed to be carried out here, giving rise to
\emp{my\_walkers} on each processor.  Third, a new function
\emp{dynamic\_load\_balancing} is called during each time step. This
function will be explained below in detail.  Fourth, unlike that the
serial counterpart could simply pass the size of its walkers to
\emp{finalize\_timestep}, the parallel implementation needs to collect
the global population size before calling \emp{finalize\_timestep}.
We remark that each local population knows its own size, but not the
global population size.  For this purpose, the
\emp{dynamic\_load\_balancing} function returns the individual local
population sizes as a \emp{numpy} array.  Last, the
\emp{collect\_subproblem\_output\_args} function from
Section~\ref{sect:motive:parallel} is used to assemble all the
individual results onto the master processor before calling the
\emp{finalize} function.

As mentioned before, parallelization of population Monte Carlo
algorithms has to take into account that the total number of walkers
changes with time.  Dynamic re-distribution of the walkers is
therefore needed to avoid work load imbalance.  The generic
\emp{dynamic\_load\_balancing} function is designed for this purpose,
where we evaluate the amount of work for each processor and, if the
work distribution is too skew, we move the excess walkers from a busy
processor to a less busy one.  The function first checks the
distribution of local population sizes. If the difference between the
smallest number of walkers and the largest number of walkers exceeds
some predefined threshold, \emp{dynamic\_load\_balancing} finds a
better population distribution and redistributes the walkers: {
\begin{quote}
\renewcommand{\baselinestretch}{0.85}\footnotesize
\begin{verbatimtab}[8]
def dynamic_load_balancing(walkers, task_time, my_rank, num_procs, \
                           send, recv, all_gather):
    walkers_per_proc = all_gather(numpy.array([len(walkers)]))
    if imbalance_rate(walkers_per_proc) > walkers.threshold_factor:
        timing_list = all_gather(numpy.array([task_time]))
        rebalanced_work = find_optimal_workload(timing_list,
                                                walkers_per_proc)
        walkers = redistribute_work(walkers,
                                    walkers_per_proc,
                                    rebalanced_work,
                                    my_rank, num_procs, send, recv)
    return walkers_per_proc
\end{verbatimtab}
\end{quote}
}

Two helper functions \emp{find\_optimal\_workload} and
\emp{redistribute\_work} are used in the above implementation.  Here,
\emp{find\_optimal\_workload} finds the optimal distribution of work,
based on how much time each local population has used.  The
\emp{redistribute\_work} function carries out the re-shuffling of
walkers.  A straightforward (but not optimal) implementation of these
functions goes as follows: {
\begin{quote}
\renewcommand{\baselinestretch}{0.85}\footnotesize
\begin{verbatimtab}[8]
def find_optimal_workload(timing_list, current_work_per_proc):
    total_work = sum(current_work_per_proc)
    C = total_work/sum(1./timing_list)
    tmp_rebalanced_work = C/timing_list
    rebalanced_work = numpy.array(tmp_rebalanced_work.tolist(),'i')
    remainders = tmp_rebalanced_work-rebalanced_work
    while sum(rebalanced_work) < total_work:
        maxarg = numpy.argmax(remainders)
        rebalanced_work[maxarg] += 1
        remainders[maxarg] = 0
    return rebalanced_work

def redistribute_work(my_walkers, work_per_proc, rebalanced_work,
                      my_rank, num_procs, send, recv):
    difference = work_per_proc-rebalanced_work
    diff_sort = numpy.argsort(difference)
    prev_rank_min = diff_sort[0]
    while sum(abs(difference)) != 0:
        diff_sort = numpy.argsort(difference)
        rank_max = diff_sort[-1]
        rank_min = diff_sort[0]
        if rank_min == prev_rank_min and rank_max != diff_sort[1]:
            rank_min = diff_sort[1]
        if my_rank==rank_max:
            send(my_walkers.cut_slice(rebalanced_work[my_rank]),\
                                      int(rank_min))
        elif my_rank==rank_min:
            my_walkers.paste_slice(recv(int(rank_max)))
        difference[rank_min] += difference[rank_max]
        difference[rank_max] = 0
        prev_rank_min = rank_min
    return my_walkers
\end{verbatimtab}
\end{quote}
}

Careful readers will notice that two particular methods,
\emp{my\_walkers.cut\_slice} and \emp{my\_walkers.paste\_slices},
provide the capability of migrating the work load between processors
in the \emp{redistribute\_work} function. These two methods have to be
programmed in class \emp{Walkers}, like the other needed methods
described earlier: \emp{move}, \emp{get\_marker}, \emp{append},
\emp{delete}, and so on.  The \emp{cut\_slice} method takes away
excess work from a local population and the \emp{paste\_slice} method
inserts additional work into a local population.  Note that the input
argument to the \emp{cut\_slice} method is an index threshold meaning
that local walkers with indices larger than that are to be taken away.
The returned slice from \emp{cut\_slice} is a picklable Python object
that can be sent and received through MPI calls.

The generic \emp{redistribute\_work} function deserves a few more
words. Among its input arguments is the ideal work distribution,
\emp{rebalanced\_work}, which is calculated by
\emp{find\_optimal\_workload}.  The \emp{redistribute\_work} function
first calculates the difference between the current distribution,
\emp{work\_per\_proc}, and the ideal distribution.  It then
iteratively moves walkers from the processor with the most work to the
processor with the least work until the difference is evened out.

This load balancing scheme is in fact independent of population Monte
Carlo algorithms.  As long as you have an algorithm repeatedly doing a
task over time and where the amount of work in the task varies over
time, this scheme can be reused.  The only requirement is that an
application-specific implementation of class \emp{Walkers}, in terms
of method names and functionality, should match with
\emp{dynamic\_load\_balancing} and \emp{redistribute\_work}.  It
should be noted that the given implementation of the latter function
is not optimal.

The algorithm of diffusion Monte Carlo, described in
Appendix~\ref{sect:dmc}, is a typical example of a population Monte
Carlo algorithm. The implementation is described in Section
\ref{sect:pdmc} and Appendix~\ref{sect:dmc}.

\subsection{Parallel Additive Schwarz Iterations}
\label{sect:3.3}

From the perspective of communication between processors,
parallelization of the Monte Carlo algorithms is relatively
easy. Parallel Markov chain Monte Carlo algorithms only require
communication in the very beginning and end, whereas parallel
population Monte Carlo algorithms only require communication at the
end of each time step. Actually, our function-centric approach to
parallelization can allow more frequent communication.  To show the
versatility of function-centric parallelization, we apply it to an
implicit method for solving partial differential equations (PDEs)
where communication is frequent between processors.

More specifically, many PDEs can be solved by an iterative process
called {\em domain decomposition}.  The idea is to divide the global
domain, in which the PDEs are to be solved, into $n$ overlapping
subdomains.  The PDEs can then be solved in parallel on the $n$
subdomains.  However, the correct boundary condition at the
\emph{internal} subdomain boundaries are not known, thus leading to an
iterative approach where one applies boundary conditions from the last
iteration, solves for the $n$ subdomain problems again, and repeats
the process until convergence of the subdomain solutions (see
e.g.~\cite{Chan:1994:DOA,Smith:1996:DPM}).  This algorithm is commonly
called additive Schwarz iteration and can successfully be applied to
many important classes of PDEs
\cite{Cai2005AWR,Cai.2005.5,Langtangen:01b}.  The great advantage of
the algorithm, especially from a software point of view, is that the
PDE solver for the global problem can be reused for each subdomain
problem.  Some additional code is needed for communicating the
solutions at the internal boundaries between the subdomains. This code
can be implemented in a generic fashion in Python, as we explain
later.

Let us first explain the additive Schwarz algorithm for solving PDEs
in more detail.  We consider some stationary PDE defined on a global
domain $\Omega$:
\begin{eqnarray}
\mathcal L(u)&=&f, \quad x\in\Omega,
\label{eqn:global-system}
\end{eqnarray}
subject to some boundary condition involving $u$ and/or its
derivatives.  Dividing $\Omega$ into a set of overlapping subdomains
$\{\Omega_s\}_{s=1}^P$, we have the restriction of
(\ref{eqn:global-system}) onto $\Omega_s$, for all $s$, as
\begin{eqnarray}
\mathcal L(u) &=&f, \quad x\in\Omega_s.
\label{eqn:sub-system}
\end{eqnarray}
The additive Schwarz method finds the global solution $u$ by an
iterative process that generates a series of approximations $u_0$,
$u_1$, $u_2$ and so on.  During iteration $k$, each subdomain computes
an improved local solution $u_{s,k}$ by locally solving
(\ref{eqn:sub-system}) for $u=u_{s,k}$ with $u_{s,k}=u_{k-1}$ as (an
artificial) boundary condition on $\Omega_s$'s non-physical internal
boundary that borders with neighboring subdomains.  All the subdomains
can {\em concurrently} carry out the local solution of
(\ref{eqn:sub-system}) within iteration $k$, thus giving rise to
parallelism.  At the end of iteration $k$, neighboring subdomains
exchange the latest local solutions in the overlapping regions to
(logically) form the global field $u_k$.  The subdomain problems
(\ref{eqn:sub-system}) are of the same type as the global problem
(\ref{eqn:global-system}), which implies the possibility of reusing an
existing serial code that was originally implemented for
(\ref{eqn:global-system}).  The additional code for exchange of local
solutions among neighbors can be implemented by generic communication
operations, independently of specific PDEs.

A generic implementation of parallel additive Schwarz iteration
algorithm can be realized as the following Python function: {
\begin{quote}
\renewcommand{\baselinestretch}{0.85}\footnotesize
\begin{verbatimtab}[8]
def additive_Schwarz_iterations(subdomain_solve, communicate, 
    set_BC, max_iter, threshold, solution, 
    convergence_test=simple_convergence_test):
    iter = 0;  not_converged = True   # init

    while not_converged and iter < max_iter:
        iter += 1
        solution_prev = solution.copy()
        set_BC(solution)
        solution = subdomain_solve()
        communicate(solution)
        not_converged = not convergence_test(\
                        solution, solution_prev, threshold)
\end{verbatimtab}
\end{quote}
}

In the above function, \emp{max\_iter} represents the maximum number
of additive Schwarz iterations allowed, and \emp{subdomain\_solve} is
a function that solves the subdomain problem of form
(\ref{eqn:sub-system}) and returns an object \emp{solution}, which is
typically a \emp{numpy} array containing the latest subdomain solution
$u_{s,k}$ on a processor (subdomain).  However, \emp{solution} may
well be a more complex object, say holding a collection of scalar
fields over computational grids, provided that (i) the object has a
\emp{copy} method, (ii) \emp{convergence\_test} and \emp{communicate}
can work with this object type, and (iii) \emp{subdomain\_solve}
returns such an object.  This flexibility in choosing \emp{solution}
reflects the major dynamic power of Python and provides yet another
illustration of the generality of the examples in this paper.

Given an existing serial code, for example in a language like Fortran
or C/C++, the \emp{subdomain\_solve} function is easily defined by
wrapping up an appropriate piece of the serial code as a Python class
(since \emp{subdomain\_solve} does not take any arguments, the
function needs a state with data structures, conveniently implemented
as class attributes as explained in Section~\ref{sect:motive:serial}).

The \emp{communicate} argument is a function for exchanging the latest
local solutions among the subdomains. After the call, the
\emp{solution} object is updated with recently computed values from
the neighboring subdomains, and contents of \emp{solution} have been
sent to the neighbors.  The \emp{communicate} function is problem
independent and can be provided by some library.  In our
implementation, the implementation is entirely in Python to take
advantage of easy programming of parallel communication in Python.
The \emp{set\_BC} argument is a function for setting boundary
conditions on a subdomain's internal boundary.  This function depends
on the actual serial code and is naturally implemented as part of the
class that provides the \emp{subdomain\_solve} function.

The \emp{convergence\_test} function is assumed to perform an
appropriate convergence test. The default generic implementation can
test
$$\max_{1\le s\le P}\frac{\|u_{s,k}-u_{s,k-1}\|^2}{\|u_{s,k}\|^2}
$$ against a prescribed threshold value.  An implementation reads
\begin{quote}
\renewcommand{\baselinestretch}{0.85}\footnotesize
\begin{verbatimtab}[8]
def simple_convergence_test(solution, solution_prev, threshold=1E-3):
    diff = solution - solution_prev
    loc_rel_change = vdot(diff,diff)/vdot(solution,solution)
    glob_rel_change = all_reduce(loc_rel_change, MAX)
    return glob_rel_change < threshold
\end{verbatimtab}
\end{quote}
We remark that \emp{all\_reduce} is a wrapper of the MPI
\emp{MPI\_Allreduce} command and \emp{vdot} computes the inner product
of two \emp{numpy} arrays.

Unlike the three-component structure described in
Sections~\ref{sect:3.1} and \ref{sect:3.2}, the main ingredients for
parallel additive Schwarz iterations are the functions of
\emp{subdomain\_solve}, \emp{communicate}, \emp{set\_BC}, and
\emp{convergence\_test}.  In other words, it is not natural to divide
the work of solving a PDE into \emp{initialize}, \emp{func}, and
\emp{finalize}.  Nevertheless, function-centric parallelization is
also here user-friendly and gives a straightforward implementation of
\emp{additive\_Schwarz\_iterations} as above.  The
\emp{convergence\_test} function shown above is clearly generic, and
so is the \emp{communicate} function in the sense that it does not
depend on the PDE. Both functions can be reused for different PDEs.
The other two functions are PDE dependent, however,
\emp{subdomain\_solve} normally wraps an existing serial code, while
the implementation of \emp{set\_BC} is typically very simple.

\section{Applications and Numerical Experiments}
\label{sect:results}

In this section we will address three real research projects involving
the three classes of algorithms covered in
Section~\ref{sect:highlevel}. The projects have utilized our
function-centric approach to parallelizing existing codes.  That is,
we had some software in Fortran, C++, and R performing the basic
computations needed in the projects.  The serial software was wrapped
in Python, adapted to components such as \emp{initialize}, \emp{func},
\emp{do\_timestep}, \emp{finalize}, \emp{subdomain\_solve},
\emp{communicate}, \emp{set\_BC}.  Parallelization was then carried
out as explained in previous sections.  An important issue to be
reported is the parallel efficiency obtained by performing the
parallelization in a Python layer that is separate from the underlying
serial scientific codes.

The Python enabled parallel codes have been tested on a Linux cluster
of 3.4 GHz Itanium2 processors, which are interconnected through
1Gbits ethernet.  The purpose is to show that the function-centric
parallelization approach is easy to use and that satisfactory parallel
performance is achievable.

\subsection{Parallel Markov Chain Monte Carlo Simulations}
\label{sect:pMCMC}

The first case is from political science and concerns estimating
legislators' ideal points by the Markov chain Monte Carlo (MCMC)
method.  For a detailed description of the mathematical problem and
the numerical method, we refer the reader to
Appendix~\ref{sect:voting}.  This application fits into the setup in
Section~\ref{sect:3.1}.  The statistical engine is provided by the
PSCL library~\cite{Jackman2006} in R~\cite{R}, for which there exists
a Python wrapper.

To use the function-centric parallelization described in
Section~\ref{sect:3.1}, we have written a Python class named
\emp{PIPE}. In addition to the constructor of the class (i.e., the
\emp{\_\_init\_\_} method), there are three methods as follows:
\begin{itemize}
\item \emp{initialize} sets up the functionality of the PSCL library
  through the Python wrapper of R (named \emp{rpy}), and prepares the
  input argument list needed for \emp{func}.
\item \emp{func} carries out the computation of each task by invoking
  appropriate functions available through \emp{rpy} (in short,
  \emp{func} is a Python wrapper to the R function \emp{ideal} from
  the PSCL library).
\item \emp{finalize} summarizes the output and generates an array in R
  format.
\end{itemize}

The resulting parallel Python program is now as short as {
\begin{quote}
\renewcommand{\baselinestretch}{0.85}\footnotesize
\begin{verbatimtab}[8]
from function_centric import parallel_solve_problem
import pypar
my_rank = pypar.rank()
num_procs = pypar.size()

from pypipe import PIPE
problem = PIPE("EP1.RData", "rcvs", "NULL", "NULL")
parallel_solve_problem(problem.initialize, problem.func, problem.finalize,
                       my_rank, num_procs, pypar.send, pypar.receive)
pypar.finalize()
\end{verbatimtab}
\end{quote}
}

The practical importance of a parallel MCMC code is that large and
computationally intensive simulations are now easily doable. More
specifically, data from the European Parliament between 1979 and
2004~\cite{Hix2005} are used for simulation.  During the five year
legislative terms, the European Parliament expanded the size of the
membership as well as the number of votes taken. (This trend has
continued since 2004.) It is hence increasingly computationally
intensive to estimate the ideal point model without reducing the
length of the Markov chain. We examined the parallel performance by
comparing the computing time for each of the five legislatures,
running the parallelized code on 32 CPUs. The results are reported in
Table~\ref{tab:results}. When comparing the results, the reader should
note that we have not made any attempts to optimize the \emp{ideal}
code (called by our \emp{func} function) for the purpose of
parallelization.  This makes it straightforward to switch to new
versions of the \emp{ideal} function. We ran 100,000 MCMC
iterations. The parallel efficiency was about 90\%.

\begin{table}[ht]
\caption{\label{tab:results}Speedup results associated with voting analysis.}
\begin{indented}
\item[]\begin{tabular}{@{}llllll}
\br
\textbf{Legislature}& Votes& Members& 1 CPU &32 CPUs& Efficiency\\
\mr
  1979 - 1984 &   810 & 548& 287m 32.560s & 10m 13.318s & 87.91$\%$\\
 1984 - 1989 &   1853& 637 & 783m 59.059s& 26m 58.702s& 91.06$\%$\\
 1989 - 1994 & 2475 & 597 & 1006m 59.258s &33m 26.140s & 94.11$\%$\\
 1994 - 1999 & 3603 & 721 & 1905m 0.930s&66m 0.068s&90.20$\%$\\
 1999 - 2004 & 5639 & 696&2898m 45.224s & 102m 7.786s &88.70$\%$\\
\br
\end{tabular}
\end{indented}
\end{table}

\subsection{Parallel Diffusion Monte Carlo Simulations}
\label{sect:pdmc}

As an example of population Monte Carlo methods, we will now look at
parallel Diffusion Monte Carlo (DMC) computations (see
Appendix~\ref{sect:dmc} for a detailed numerical description), which
is used here to simulate Bose-Einstein condensation.  We recall from
Section~\ref{sect:3.2} that dynamic load balancing is needed in
connection with the parallelization, and can be provided by the
generic \emp{dynamic\_load\_balancing} function.  To utilize the
parallel time integration function \emp{parallel\_time\_integration}
from Section~\ref{sect:3.2}, we need to program a parallel version of
the \emp{initialize} function. The \emp{do\_timestep} function from
Section~\ref{sect:3.2} can be used as is.  {
\begin{quote}
\renewcommand{\baselinestretch}{0.85}\footnotesize
\begin{verbatimtab}[8]
def initialize(my_rank, num_procs):
    nwalkers = 1000
    nspacedim = 3
    stepsize = 0.1
    timesteps = 200
    walkers_per_proc = simple_partitioning(nwalkers, num_procs)
    my_walkers = Walkers(walkers_per_proc[my_rank], nspacedim, stepsize)        
    my_walkers.threshold_factor = 1.1
    return my_walkers, timesteps
\end{verbatimtab}
\end{quote}
} This \emp{initialize} function is quite similar to its serial
counter part. As noted in Section~\ref{sect:3.2}, it takes as input
\emp{my\_rank} and \emp{num\_procs}. The \emp{simple\_partitioning}
function is called to partition the walker population. A
\emp{my\_walkers} object is assigned to each processor, and a
threshold factor is prescribed to determine when load balancing is
needed.

Together with the \emp{parallel\_time\_integration} function from
Section~\ref{sect:3.2}, the above \emp{initialize} function is the
minimum programming effort needed to parallelize a serial population
Monte Carlo code. For the particular case of our parallel Diffusion
Monte Carlo implementation, we also need to know the global number of
walkers in every timestep to be able to estimate its observables
globally. Moreover, the load balancing scheme needs the time usage of
each processor during each time step.

A class with name \emp{Walkers} needs to be implemented to match with
the implementations of \emp{parallel\_time\_integration},
\emp{dynamic\_load\_balancing}, and the above \emp{initialize}
function.  The essential work is to provide a set of methods with
already decided names (see Section~\ref{sect:3.2}), such as
\emp{move}, \emp{append}, \emp{delete}, \emp{finalize\_timestep},
\emp{cut\_slice}, and \emp{paste\_slice}.  A concrete example of the
\emp{Walkers} class is described with more details in
Appendix~\ref{sect:dmc}.

We report in Table~\ref{tab:results2} the timing results of a number
of parallel DMC computations.  The parallel efficiency was about 85\%.
We increased the total number of walkers when more processors were
used in the simulation, such that the number of walkers assigned to
each processor remained as 200. Such a use of parallel computers for
DMC simulations mimics the everlasting wish of quantum physicists to
do larger computations as soon as more computing resource becomes
available.  Note that in this scaled scalability test, good speedup
performance is indicated by an almost constant time usage independent
of the number of processors.

\begin{table}[ht]
\caption{\label{tab:results2}Timing results of the parallel DMC
  simulations where each processor is constantly assigned with 200
  walkers, all moved in 5000 time steps.}
\begin{indented}
\item[]\begin{tabular}{@{}lll}
\br
CPUs&Time&Efficiency\\
\mr
1&37m10.389s &N/A\\
5&42m32.359s  &87.39\%\\
10&42m00.734s &88.48\%\\
20&42m29.945s &87.47\%\\
30&42m33.895s &87.33\%\\
40&43m30.092s &85.45\%\\
50&43m39.159s &85.16\%\\
\br
\end{tabular}
\end{indented}
\end{table}

\subsection{Parallel Boussinesq Simulations}

Simulating the propagation of ocean waves is the target of the our
third and final concrete case.  The reader is referred to
Appendix~\ref{sect:add_sch} for the mathematical model and the
numerical method.  The involved equations can be solved in parallel by
the additive Schwarz algorithm of Section~\ref{sect:3.3}.

Our starting point for parallelization is a 25 years old legacy
Fortran 77 code consisting of a set of subroutines. More specifically,
the most important subroutines are \emp{KONTIT} and \emp{BERIT}, which
target the two semi-discretized equations (\ref{eqn:dist-continu}) and
(\ref{eqn:dist-bernoulli}) of the mathematical model (see
Appendix~\ref{sect:add_sch}).  These two Fortran 77 subroutines
contain intricate algorithms with nested layers of do-loops, which are
considered to be very difficult to parallelize by directly inserting
MPI calls in the Fortran code.  Performing the parallelization outside
the Fortran code is therefore much more convenient. Using the proposed
framework in the present paper the parallelization is a technically
quite straightforward task.

The subdomain solver consists of calls to the subroutines \emp{KONTIT}
and \emp{BERIT}. The implementation of the Python function
\emp{subdomain\_solve} (see Section~\ref{sect:3.3}) requires a Python
interface to \emp{KONTIT} and \emp{BERIT}, which can easily be
produced by the F2PY software.  Since a subdomain solver needs to set
artificial boundary conditions at non-physical boundaries, we have
programmed two light-weight wrapper subroutines in Fortran,
\emp{WKONTIT} and \emp{WBERIT}, which handles the boundary conditions
before invoking \emp{KONTIT} and \emp{BERIT}.  We then apply F2PY to
make \emp{WKONTIT} and \emp{WBERIT} callable from Python. Since the
Fortran subroutines have lots of input data in long argument lists and
\emp{subdomain\_solve} takes no arguments, we have created a class
where the Fortran input variables are stored as class attributes:
\begin{quote}
{
\renewcommand{\baselinestretch}{0.85}\footnotesize
\begin{verbatimtab}[8]
import f77  # extension module for the Fortran code

class SubdomainSolver:
    def __init__(self, ...):
        # set input arguments to the Fortran subroutines as class attributes
        # (nbit,F,YW,H,QY,WRK,dx,dy,dt,kit,ik,gg,alpha,eps,
        # lower_x_neigh,upper_x_neigh,lower_y_neigh,upper_y_neigh)

    def continuity(self):
        self.Y, self.nbit = f77.WKONTIT(\
          self.F, self.Y, self.YW, self.H, self.QY, self.WRK,
          self.dx, self.dy, self.dt, self.kit, self.ik,
          self.gg, self.alpha, self.eps, self.nbit,
          self.lower_x_neigh, self.upper_x_neigh,
          self.lower_y_neigh, self.upper_y_neigh)

    def bernoulli(self):
        # similar to the continuity method

sd = SubdomainSolver(...)
subdomain_solve1 = sd.continuity
subdomain_solve2 = sd.bernoulli
\end{verbatimtab}
}
\end{quote}

Note that since there are two PDEs (\ref{eqn:dist-continu}) and
(\ref{eqn:dist-bernoulli}), we have created two functions:
\emp{subdomain\_solve1} and \emp{subdomain\_solve2}.  The main
computation of the resulting parallel program is in the following
while loop:
\begin{quote}
{
\renewcommand{\baselinestretch}{0.85}\footnotesize
\begin{verbatimtab}[8]
t = 0
while t < t_stop:
    t = t+dt
    additive_Schwarz_iterations(subdomain_solve1, communicate, 
                                set_BC, 10, 1E-3, sd.Y)
    additive_Schwarz_iterations(subdomain_solve2, communicate, 
                                set_BC, 10, 1E-3, sd.F)
\end{verbatimtab}
}
\end{quote}
The \emp{additive\_Schwarz\_iterations} function from
Section~\ref{sect:3.3} can be placed in a reusable module. The
\emp{communicate} function is borrowed from a Python library for mesh
partitioning and inter-subdomain communication. The \emp{set\_BC}
function actually does not do anything for this particular
application.

Speedup results are reported in Table~\ref{tab:results3}, for which
the global solution mesh was fixed at $1000\times1000$, and the number
of time steps was 40.  The results show that we can handle a quite
complicated mathematical problem in a black-box Fortran code with our
suggested simple framework and obtain a remarkable good speedup, with
just a trivial extension of the Fortran code.

\begin{table}[ht]
\caption{\label{tab:results3}The speedup results of the Python enabled
  parallel Boussinesq simulations.}
\begin{indented}
\item[]\begin{tabular}{@{}llll}
\br
CPUs & Time & Speedup & Efficiency\\
\mr
1 & 166.66s & N/A & N/A\\
2 & 83.61s & 1.99 & 99.67\%\\
4 & 44.45s & 3.75 & 93.73\%\\
8 & 20.16s & 8.27 & 103.33\%\\
16 & 11.43s & 14.58 & 91.13\%\\
\br
\end{tabular}
\end{indented}
\end{table}

\section{Conclusion}
\label{sect:conclusion}

We have shown how serial scientific codes written in various common
languages, including Fortran, C, C++, and Python, can be parallelized
in a separate, small software unit written in Python.  The advantage
of such an approach is twofold. First, the existing, often
complicated, scientific high-performance code remains (almost)
unchanged. Second, the parallel algorithm and its inter-processor
communication are conveniently implemented in high-level Python code.

This approach to parallelization has been implemented in a software
framework where the programmer needs to implement a few Python
functions for carrying out the key steps in the solution approach. For
example, our first application involves doing a set of independent
tasks in parallel, where a small Python framework deals with the
parallelism and demands the user to supply three functions:
\emp{initialize} for preparing input data to the mathematical model,
\emp{func} for calling up the serial scientific code, and
\emp{finalize} for processing the computational results. Some more
functions must be supplied in more complicated problems where the
algorithm evolves in time, with a need for dynamic load balancing and
more parallel communication.

Our simple software frameworks outlined in this paper are applicable
to many different scientific areas, and we have described some common
classes of problems: parameter investigation of a mathematical model,
standard Monte Carlo simulation, Monte Carlo simulation with need for
dynamic load balancing, and numerical solution of partial differential
equations.  In each of these cases, we have outlined fairly detailed
Python code such that most technical details of the parallel
implementations are documented. This may ease the migration of the
ideas to new classes of problems beyond the scope of this paper.

In particular, the shown frameworks have been used to parallelize
three real scientific problems taken from our research. The problems
concern Markov Chain Monte Carlo models for voting behavior in
political science, Diffusion Monte Carlo methods for Bose-Einstein
condensation in quantum mechanics, and additive Schwarz and finite
difference methods for simulating ocean waves by a system of partial
differential equations.  The results of our investigations of the
parallel efficiency are very encouraging: In all these real science
problems, parallelizing serial codes in the proposed Python framework
gives almost optimal speedup results, showing that there arises no
significant loss due to using Python and performing the
parallelization ``outside'' the serial codes.

As a conclusion, we believe that the ideas and code samples from this
paper can simplify parallelization of serial codes greatly, without
significant loss of computational efficiency. This is good news for
scientists who are non-experts in parallel programming but want to
parallelize their serial codes with as small efforts as possible.

\newpage
\appendix
\section{Voting in Legislatures}
\label{sect:voting}

In the spatial model of politics, both actors' preferences over
policies (ideal points) and policy alternatives are arranged
geometrically in a low-dimensional Euclidean space. An actor receives
the highest possible utility if a policy is located at her ideal
point; she gains or loses utility as the policy moves towards or away
from her ideal point~\cite{Hinich1997}.  We adopt the Bayesian
approach proposed by Clinton, Jackman and Rivers~\cite{Clinton2004}.
Assume there are $n$ legislators who vote on $m$ proposals.  On each
vote $j=1,\ldots,m$, legislator $i=1,\ldots,n$ chooses between a "Yea"
position $\zeta_{j}$ and a "Nay" position $\psi_{j}$ located in the
policy-space $\mathbb{R}^{d}$, where $d$ is the number of
dimensions. Then, we have $y_{ij}=1$ if legislator $i$ votes "Yea" on
roll call $j$, and $y_{ij}=0$ if she votes "Nay". The model assumes
quadratic utility functions. The ideal point of legislator $i$ is
\textbf{$x_{i}$} $\in \mathbb{R^{d}}$, while $\eta_{ij}$ and
$\upsilon_{ij}$ are stochastic elements whose distribution is jointly
normal. The variance of the stochastic elements is
$(\eta_{ij}-\upsilon_{ij})=\sigma^{2}_{j}$. Denote the Euclidean norm
by $\| \cdot \|$, utility maximising implies that legislator $i$ votes
"Yea" on vote $j$ if
\begin{eqnarray}
U_{i}(\zeta_{j})= - \| x_{i} - \zeta_{j} \|^{2} +
\eta_{ij} > U_{i}(\psi_{j})= - \| x_{i} - \psi_{j}
\|^{2} + \upsilon_{ij}
\end{eqnarray}
and "Nay" otherwise. Clinton, Jackman and Rivers~\cite{Clinton2004}
show that the model can be understood as a hierarchical probit model:
\begin{eqnarray}
P(y_{ij}=1)=\Phi(\beta^{\prime}_{j}\mathbf{x}_{i}
-\alpha_{j}),
\end{eqnarray}
where $\beta_{j}$=$2(\zeta_{j}-\psi_{j})/\sigma_{j}$,
$\alpha_{j}$=$(\zeta^{\prime}_{j}\zeta_{j}-\psi^{\prime}_{j}\psi_{j})/\sigma_{j}$,
$\Phi(\cdot)$ is the standard normal function, \textbf{$\beta_{j}$} is
the midpoint between the "Yea" and "Nay" positions on proposal $j$,
and $\mathbf{x}_{i}$ is the legislator's ideal point. The direction of
\textbf{$\alpha_{j}$} indicates the location of the status quo
relative to the proposal. If \textbf{$\alpha_{j}$} is positive, the
new proposal is located higher on the dimension than the status
quo. If \textbf{$\alpha_{j}$} is negative, the new proposal is located
lower on the dimension than the status quo.

\paragraph{The MCMC Algorithm.}
In the Markov Chain Monte Carlo (MCMC) algorithm for the statistical
analysis of voting behavior~\cite{Clinton2004}, the difference between
utilities of the alternatives on the $j$th vote for the $i$th
legislator is given by $y_{ij}^*=\beta_j\vec x_i - \alpha_j +
\epsilon_{ij}$, where $\beta_j$ and $\alpha_j$ are model parameters,
$\vec x_i$ is a vector of regression coefficients and $\epsilon_{ij}$
are standard normal errors. If we know $\beta_j$ and $\alpha_j$, $\vec
x_i$ can be imputed from the regression of $y_{ij}^*+\alpha_j$ on
$\beta_j$ using the $m$ votes of legislator $i$ and vice versa. If we
know $\vec x_i$, we can use the votes of the $n$ legislators on roll
call $j$ to find $\beta_j$ and $\alpha_j$. Given $\vec x_i$, $\beta_j$
and $\alpha_j$ (either from priors or from the previous iteration), we
can find $y_{ij}^*$ by drawing $\epsilon_{ij}$ randomly from a normal
distribution subject to the constraints implied by the actual votes,
i.e., if $y_{ij}=0$, $y_{ij}^*<0$ and if $y_{ij}=1$, $y_{ij}^*>0$.

The goal is to compute the joint posterior density for all model
parameters $\beta_j$ and $\alpha_j$, $j=1,\ldots, m$ and all
coefficient vectors $x_i$, $i=1,\ldots, n$. The MCMC algorithm forms a
Markov chain to explore as much as possible of this joint density,
i.e., letting $t$ index an MCMC iteration,
\begin{enumerate}
\item find $y_{ij}^{*,t}$ from $\vec x_i^{t-1}$, $\beta_j^{t-1}$ and
  $\alpha_j^{t-1}$,
\item sample $\beta_j^{t}$ and $\alpha_j^{t}$ using $\vec x_i^{t-1}$
  and $y_{ij}^{*,t}$,
\item find $\vec x_i^{t}$ from $\beta_j^{t}$, $\alpha_j^{t}$ and
  $y_{ij}^{*,t}$.
\end{enumerate}
This process must then be repeated until convergence, i.e., that the
samples have moved away from the priors to the neighborhood of the
posterior mode before samples are drawn.

Clinton, Jackman and Rivers~\cite[p.~369]{Clinton2004} find that the
computation time is increasing in $nmT$, where $n$ is the number of
legislators, $m$ is the number of votes and $T$ is the number of MCMC
iterations. Although they argue that very long runs are normally not
necessary, they nevertheless advise long runs to ensure that the MCMC
algorithm has converged. It is increasingly time-consuming to estimate
the model on on a standard desktop computer as the size of the
legislature and the number of votes increase.

\section{Bose-Einstein Condensation}
\label{sect:dmc}

The famous experiment of Anderson et al.~\cite{anderson} was about
cooling $4\times 10^6$ $^{87}$Rb down to temperatures in the order of
$100\, nK$ for observing Bose-Einstein condensation in the dilute
gas. To model this fascinating experiment in the framework of Quantum
Monte Carlo, so that numerical simulations can be extended beyond the
physical experiments, we may use the governing Schr\"odinger equation:

\begin{equation}
i\hbar\frac{\partial}{\partial t}\Psi(\vec R,t) = H\Psi(\vec R,t).
\label{eq:SE_time}
\end{equation}

The most important parts of the mathematical model are a Hamiltonian
$H$ and a wave function $\Psi$, see~\cite{nilsen2005}.  The
Hamiltonian for $N$ trapped interacting atoms is given by
\begin{equation}
H=-\frac{\hbar^2}{2m}\sum_{i=1}^{N}\nabla_i^2 + \sum_{i=1}^{N}
V_{ext}(\vec r_i) +\sum_{i<j}^{N} V_{int}(|\vec r_i - \vec r_j|).
\label{eq:Hamiltonian}
\end{equation}
The external potential $V_{ext}$ corresponds to the trap used to
confine the $^{87}$Rb atoms, and was in the experiment in the order of
$\vec r^2$.  The two-body interaction $V_{int}(|\vec r_i - \vec r_j|)$
can be easily described by a hard-core potential of radius $a$ in a
dilute gas. We have however, for the sake of simplicity, neglected
these interactions in our example implementation of class
\emp{Walkers}.

\paragraph{The Method of Diffusion Monte Carlo.}

In the Diffusion Monte Carlo (DMC) method~\cite{guardiola}, the
Schr\"odinger equation is solved in imaginary time,
\begin{equation}
-\frac{\partial d\psi(\vec R,t)}{\partial t} = [\oper H-E]\psi(\vec
R,t).
\label{eq:SL_imaginary_time}
\end{equation}
The formal solution of \refeq{eq:SL_imaginary_time} is
\begin{equation}
\psi(\vec R,t) = e^{-[H-E]t}\psi(\vec R,0),
\end{equation}
where $e^{[-(H-E)t]}$ is called the {\it Green's function}, and
\matenv E is a convenient energy shift.

The wave function $\psi(\vec R,t)$ in DMC is represented by a set of
random vectors \matenv{\{R_1,R_2,\ldots,R_M\}}, in such a form that
the time evolution of the wave function is actually represented by the
evolution of a set of walkers. This feature gives rise to task
parallelism.  The wave function is positive definite everywhere, as it
happens with the ground state of a bosonic system, so it may be
considered as a probability distribution function.

The DMC method involves Monte Carlo integration of the Green's
function by every walker.  The time evolution is done in small time
steps $\tau$, using the following approximate form of the Green's
function:
\begin{equation}
e^{-[H-E]t}=\prod_{i=1}^ne^{-[H-E]\tau},
\end{equation}
where \matenv{\tau=t/n}.  Assume that an arbitrary starting state can
be expanded in the basis of stationary,
\begin{equation}
\psi(\vec R,0)=\sum_\nu C_\nu\phi_\nu(\vec R),
\end{equation}
we have
\begin{equation}
\psi(\vec R,t)=\sum_\nu e^{-[E_\nu-E]t}C_\nu\phi_\nu(\vec R),
\label{eq:imaginary_time_evolution}
\end{equation}
in such a way that the lowest energy components will have the largest
amplitudes after a long elapsed time, and in the limit of
\matenv{t\to\infty} the most important amplitude will correspond to
the ground state (if \matenv{C_0\neq0})\footnote{This can easily be
  seen by replacing \matenv E with the ground state energy
  \matenv{E_0} in \refeq{eq:imaginary_time_evolution}. As \matenv{E_0}
  is the lowest energy, we will get
  \matenv{\lim_{t\to\infty}\sum_\nu\exp[-(E_\nu-E_0)t] \phi_\nu =
    C_0\phi_0}.}.

The Green's function is approximated by splitting it up in a
diffusional part,
\begin{equation}
G_{D} = (4\pi D\tau)^{-3N/2}\exp\{-(\vec R'-\vec R)^2/4D\tau\},
\label{eq:gdiff}
\end{equation}
which has the form of a Gaussian and a branching part,
\begin{equation}
G_{B} = \exp\{-((V(\vec R)+V(\vec R'))/2 - E_T)\tau\}.
\label{eq:gbranch}
\end{equation}
While diffusion is taken care of by a Gaussian random distribution,
the branching is simulated by creation and destruction of walkers with
a probability $G_{B}$.
The idea of DMC computation is quite simple; once we have found an
appropriate approximation of the short-time Green's function and
determined a starting state, the computation consists in representing
the starting state by a collection of walkers and letting them
independently evolve in time. That is, we keep updating the walker
population, until a large enough time when all other states than the
ground state are negligible.

~\newline
\begin{algorithm}
\rm
\begin{center}
  {Diffusion Monte Carlo}
  \label{alg:dmc}\\
    \fbox{\parbox{0.9\textwidth}{
  \begin{tabbing}
\hspace*{0.5cm}\= \hspace{0.5cm} \= \hspace{0.5cm} \=
\hspace{0.5cm} \= \hspace{0.5cm} \= \kill
\textbf{for} \texttt{step} \textbf{in range(} \texttt{0}, \texttt{timesteps} \textbf{)\,:}\\
\>\textbf{for} \texttt{$i$} \textbf{in range(} \texttt{0}, $N_{\mathrm{walkers}}$ \textbf{)\,:}\\
  \>\>Diffusion;\\
  \>\>\>propose move \matenv{\vec R'=\vec R + \xi}\\
   \>\>Branching;\\
   \>\>\>calculate replication factor $n$:\\
   \>\>\>\>\matenv{n=int(\exp\{-((V(\vec R)+V(\vec R'))/2-E_T)\tau\})}\\
   \>\>\>\textbf{if} \matenv{n=0}\textbf{\,:}\\
   \>\>\>\>mark walker as dying\\
   \>\>\>\textbf{if} \matenv{n>0}\textbf{\,:}\\
   \>\>\>\>mark walker to make \matenv{n-1} clones\\
   \>\>Remove dead walkers and make new clones;\\
   \>\>Update walker population $N_{\mathrm{walkers}}$ and adjust trial energy;\\
   \>\>Sample contributions to observable.
  \end{tabbing}
}}
\end{center}
\end{algorithm}

\paragraph{The Implementation.}

In Algorithm~\ref{alg:dmc} we summarize the DMC algorithm
corresponding to \refeq{eq:gdiff}-\refeq{eq:gbranch}.  In the
algorithm $\xi$ is a Gaussian with zero mean and a variance of
$2D\tau$ corresponding to \refeq{eq:gdiff}.  The deleting and cloning
of walkers are, as mentioned in Section~\ref{sect:3.2}, performed by
the \emp{do\_timestep} function, repeated here for clarity:
\begin{quote}
{
\renewcommand{\baselinestretch}{0.85}\footnotesize
\begin{verbatimtab}[8]
def do_timestep(walkers):
    walkers.move()
    for walker in range(len(walkers)):
        if walkers.get_marker(walker) == 0:
            walkers.delete(walker)
        elif walkers.get_marker(walker) > 1:
            walkers.append(walker, walkers.get_marker(walker)-1)        
    return walkers.sample_observables()
\end{verbatimtab}
}
\end{quote}

The main computational work of the DMC algorithm at each time step is
implemented in the \emp{move} function inside class \emp{Walkers},
together with a helper function \emp{branching}:
\begin{quote}
{
\renewcommand{\baselinestretch}{0.85}\footnotesize
\begin{verbatimtab}[8]
class Walkers:
    ...

    def branching(self, new_positions):
        old_potential = potential(self.positions)
        new_potential = potential(new_positions)
        branch = numpy.exp(-(0.5 * (old_potential + new_potential)
                         - self.adjust_branching) * self.stepsize)
        self.markers = numpy.array(branch+
                      numpy.random.uniform(0,1,branch.shape), 'i')
        
    def move(self):
        displacements = numpy.random.normal(0, 2*self.stepsize, 
                                             self.positions.shape)
        new_positions = self.positions+displacements
        self.branching(new_positions)
        self.positions = new_positions
    ...

\end{verbatimtab}
}
\end{quote}
The \emp{move} function first generates a set of Gaussian (normal)
distributed random numbers, corresponding to \refeq{eq:gdiff}. Next,
it calls the \emp{branching} function, which calculates a potential
$V(\vec r)=\vec r^2$ for the old and the new positions\footnote{In a
  more optimized implementation, the old potential would have been
  stored from the previous move and not calculated every time.}. These
potentials are used to calculate $G_B$ following \refeq{eq:gbranch}
and create an integer array \emp{self.markers} with its average value
equal to $G_B$ (stored in the \emp{branch} variable).  This array is
of the same length as the number of walkers (stored in
\emp{self.positions}) and marks the walkers as dying or clone-able.

It is worth noticing that if the new potential of a walker is much
higher than that in the previous time step (i.e., the walker is far
from the center of the trap), the value of \emp{branch} will be close
to 0 and the walker will be deleted. However, if the new potential is
much lower (i.e. closer to the center of the trap), \emp{branch} will
be greater than 1 and the walker will be cloned. As long as the
two-body interaction is ignored, the walkers will only be encouraged
to move towards the center of the trap, thus yielding a lower energy
than seen in real experiments.

\section{Ocean Wave Propagation}
\label{sect:add_sch}

The following two PDEs, normally termed as the Boussinesq water wave
equations~\cite{Wu:1982}, can be used to model wave propagation:
\begin{eqnarray}
\frac{\partial \eta}{\partial t}+\nabla\cdot
\left((H+\alpha\eta)\nabla \phi+\epsilon H
\left(\frac{1}{6}\frac{\partial \eta}{\partial t}
-\frac{1}{3}\nabla H\cdot \nabla \phi\right)\nabla H\right)&=&0, \label{eqn:continu}\\
\frac{\partial\phi}{\partial t}+\frac{\alpha}{2}\nabla\phi\cdot
\nabla\phi+\eta -\frac{\epsilon}{2} H\nabla\cdot\left(
H\nabla \frac{\partial\phi}{\partial t}\right)+\frac{\epsilon}{6}
H^2\nabla^2\frac{\partial\phi}{\partial t}&=&0.
\label{eqn:bernoulli}
\end{eqnarray}
The primary unknowns of (\ref{eqn:continu})-(\ref{eqn:bernoulli}) are
the water surface elevation $\eta(x,y,t)$ and the depth-averaged
velocity potential $\phi(x,y,t)$.  The symbol $H$ denotes the water
depth as a function of $(x,y)$.  The advantage of the above Boussinesq
wave model, in comparison with the standard shallow water equations,
is its capability of modeling waves that are weakly dispersive
($\epsilon>0$) and/or weakly nonlinear ($\alpha>0$),
see~\cite{Pedersen:99b}.  Therefore, the Boussinesq water wave
equations are particularly adequate for simulating ocean wave
propagation over long distances and large water depths.

Discretization of the Boussinesq water wave equations
(\ref{eqn:continu})-(\ref{eqn:bernoulli}) normally starts with a
temporal discretization as follows:
\begin{eqnarray}
\frac{\eta^{\ell}-\eta^{\ell-1}}{\Delta t} +
\nabla\cdot\biggl(\left
(H+\alpha\frac{\eta^{\ell-1}+\eta^\ell}{2}\right)\nabla\phi^{\ell-1}
+ && \nonumber\\
\epsilon H\left(\frac{1}{6}\frac{\eta^{\ell}-\eta^{\ell-1}}{\Delta t}
-\frac{1}{3}\nabla H\cdot\nabla\phi^{\ell-1}\right)\nabla H\biggr)
&=& 0,\label{eqn:dist-continu}\\
\frac{\phi^{\ell}-\phi^{\ell-1}}{\Delta t}+\frac{\alpha}{2}
\nabla\phi^{\ell-1}\cdot\nabla\phi^{\ell-1}+\eta^{\ell}
-&& \nonumber\\
\frac{\epsilon}{2} H\nabla \cdot \left(H\nabla\left(\frac
{\phi^{\ell}-\phi^{\ell-1}}{\Delta t}\right)\right)+\frac{\epsilon}{6}
 H^2\nabla^2\left(\frac{\phi^{\ell}-\phi^{\ell-1}}{\Delta t}\right)&=&0,
\label{eqn:dist-bernoulli}
\end{eqnarray}
where we use $\ell$ to denote the time level, and $\Delta t$ denotes
the time step size. The basic idea of computation at each time step is
to first compute $\eta^{\ell}$ based on $\eta^{\ell-1}$ and
$\phi^{\ell-1}$ from the previous time step, and then compute
$\phi^{\ell}$ using the new $\eta^{\ell}$ and the old $\phi^{\ell-1}$.
To carry out the actual numerical computation, we need a spatial
discretization of (\ref{eqn:dist-continu})-(\ref{eqn:dist-bernoulli}),
using e.g.~finite differences or finite elements, so we end up with
two systems of linear equations that need to be solved during each
time step.

\section*{References}
\bibliography{parapython_iop}
\bibliographystyle{unsrt}

\end{document}